\documentstyle[12pt]{article}
\newcommand{\be}{\begin{equation}}
\newcommand{\ee}{\end{equation}}
\newcommand{\bea}{\begin{eqnarray}}
\newcommand{\eea}{\end{eqnarray}}
\newcommand{\bean}{\begin{eqnarray*}}
\newcommand{\eean}{\end{eqnarray*}}
\newcommand{\ba}{\begin{array}{l}}
\newcommand{\ea}{\end{array}}
\newcommand{\bb}{\begin{thebibliography}{99}}
\newcommand{\eb}{\end{thebibliography}}
\newcommand{\ci}[1]{\cite{#1}}
\newcommand{\lab}[1]{\label{#1}}
\newcommand{\re}[1]{(\ref{#1})}

\newcommand{\Ds}{\displaystyle}

\begin{document}
\Large
\thispagestyle{empty}
\vskip 3cm
\begin{center}
{\bf RENORMALIZATION CONDITIONS \\
AND NON-DIAGRAMMATIC APPROACH TO RENORMALIZATIONS}
\end{center}
\vskip 1cm
\large
\centerline{B.A.FAIZULLAEV}
\centerline{ Theoretical Physics Department, Tashkent State University }
\centerline{ Tashkent 700095, Uzbekistan }
\centerline{    and }
\centerline{S.A.GARNOV }
\centerline{ Institute of Applied Physics, Tashkent State University }
\centerline{    Tashkent 700095, Uzbekistan}

\normalsize

\begin{abstract}
The representation of the bare parameters of Lagrangian in terms of
total vertex Green's functions is used to obtain the general form of
renormalization conditions. In the framework of our approach
renormalizations can be carried out without treatment to Feynman diagrams.

\end{abstract}

\newpage

\section{Introduction}

The representation of the bare parameters of Lagrangian in terms of Green's
functions is usefull for investigation of a great variety of subjects
in the quantum field theory and statistical physics. For example, this
approach was applied to the theory of self-generating interactions \ci{edd},
critical phenomena \ci{da}, study of the second (third, fourth etc.) Legendre
transformation \ci{vas} and so on. We want to show that the representation
like this may be used successfully in the renormalization theory,
specifically, to obtain the renormalization conditions.

The renormalization is the redefinition of the bare parameters of Lagrangian
through inserting the infinities connected with the loop integrals into
the bare parameters \ci{del}. But to divide each infinite integral into the
finite and infinite parts we must carry out the subtraction procudure which
has ambiguities owing to the different choises of the subtraction point
(or the mass parameter $\mu$ in the dimensional regularization). Therefore
we have to impose some renormalization conditions. Usually these conditions
are postulated on the strenght of some general considerations \ci{iim},
\ci{vsh}, \ci{iz}.
We propose a method which enables us to get these conditions. Besides,
as it will be shown, within our formulation we can deal with usual
(ultraviolet) infinities without treatment to Feynmann diagrams.

The purpose of this article is to look into the old approach - the
representation of the bare parameters in terms of Green's functions - in
the light of the renormalizations and generalise it in such a way that
the bare parameters can be expressed consistently in terms of renormalized
vertex functions $\Gamma^{(n)}$ which themselwes depend on a renormalized
mass and coupling constant. This programm leads to the usual renormalization
conditions automatically.

The article is organized as follows: in Sec.2 we introduce a wethod which
allows us to represent the bare mass and coupling constant in terms of
the total vertex Green's functions and obtain the corresponding expressions.
In Sec.3 the renormalization conditions are derived in the most general form.
We shall discuss them thoroughly at the second order of $\hbar$ and show
how to apply our approach to higher orders.

\section{Bare mass and coupling constant
in terms of the vertex functions}

For definiteness we shall consider scalar fields.
Let us introduce a generating functional of connected Green's functions
by the path integral expression
\be
\exp\left(\frac{i}{\hbar}W[J]\right)=N^{-1}_{0}
\int D\varphi \exp\left(\frac{i}{\hbar}S[\varphi]+
\frac{i}{\hbar}\int dx J(x)\varphi(x)\right), \lab{e1}
\ee
where
\be N_{0}=\int D\varphi\exp\left(\frac{i}{\hbar}S[\varphi]\right) .\lab{e2}
\lab{e} \ee

The generating functional $\Gamma[\Phi]$ of the vertex functions is
determined as
\be \Gamma[\Phi]=W[J]-\int dx J(x)\Phi(x),
\lab{e3} \ee
\be \Phi (x)=\frac{\delta W[J]}{\delta J(x)}.\lab{e4} \ee

The expression \re{e3} is the functional Legendre transformation which
introduce new functional argument $\Phi$ instead of the functional argument
$J$. It follows from \re{e3} that
\be \frac{\delta \Gamma[\Phi]}{\delta \Phi(x)}=-J(x).\lab{e5} \ee

We can expand $\Gamma[\Phi]$ in the Taylor series
\be  \Gamma[\Phi]=\sum\frac{1}{n!}\int dx_1\cdots dx_n
\Gamma^{(n)}(x_1,\cdots,x_n)\Phi(x_1)\cdots\Phi(x_n), \lab{e6}
 \ee
where
\be \Gamma^{(n)}(x_1,\cdots,x_n)=\left.\frac{\delta^n\Gamma[\Phi]}{\delta
\Phi(x_1)
\cdots\delta\Phi(x_1)}\right|_{\Phi(x_1)=\cdots=\Phi(x_n)=0}
\lab{e7} \ee
are the vertex functions.

In momentum space $\Gamma^{(n)}$ is presented as
\be\ba\Ds \Gamma^{(n)}(x_1,\cdots,x_n)=\int \frac{dk_1}{(2\pi)^4}\cdots
\frac{dk_n}{(2\pi)^4} (2\pi)^4\delta(k_1+\cdots +k_n)\times   \\
\Ds\times\exp i(k_1 x_1+\cdots+k_n x_n)\Gamma^{(n)}(k_1,\cdots,k_n).\lab{e8}
\ea  \ee
Noting that the integral \re{e1} is the functional Fourier transformation
one can write the formula for the inverse transformation
\be \exp\left(\frac{i}{\hbar}S[\varphi]\right)= N
\int D(\frac{1}{\hbar}J)\ exp\left(\frac{i}{\hbar}W[J]-
\frac{i}{\hbar}\int dx J(x)\varphi(x)\right),\lab{e9}
\ee
where
$ N=(2\pi)^{-\nu} N_{0} $
($\nu$ is the dimensionality of the space $\varphi$).

Using \re{e5} and replacing of $J$ by $\Phi$
we get
\be\ba\Ds \exp\left(\frac{i}{\hbar}S[\varphi]\right)= N\int D\Phi
 \det \left[-\frac{1}{\hbar}
\frac{\delta ^2 \Gamma}{\delta\Phi ^2}\right]\times  \\
\Ds\times\exp\left(\frac{i}{\hbar}\Gamma[\Phi]-\frac{i}{\hbar}\int dx
\frac{\delta\Gamma}{\delta\Phi (x)}(\Phi (x)-\varphi (x))\right).
\lab{e10} \ea \ee

It is convenient for us to introduce the following notation
\be\ba\Ds F[\Phi,\varphi]=\det\left[-\frac{1}{\hbar}
\frac{\delta ^2 \Gamma}{\delta\Phi ^2}\right]\times  \\
\Ds\times\exp\left(\frac{i}{\hbar}\Gamma[\Phi]-\frac{i}{\hbar}\int dx
\frac{\delta\Gamma}{\delta\Phi (x)}(\Phi (x)-\varphi (x))\right).
\lab{e11} \ea\ee

Let us differentiate the both sides of \re{e10} with respect to $\varphi(x)$
\be \frac{\delta S}{\delta \varphi (x)}
\exp\left(\frac{i}{\hbar}S[\varphi]\right)= N\int D \Phi
\frac{\delta\Gamma}{\delta\Phi (x)}F[\Phi,\varphi].\lab{e12} \ee

Now let us expand $\delta\Gamma / \delta\Phi$ in the integrand in the
Taylor series
\be\ba\Ds
\frac{\delta S}{\delta\varphi (x)}
\exp\left(\frac{i}{\hbar}S[\varphi]\right)=\\
\Ds=N\Gamma^{(1)} (x)
\int D \Phi F[\Phi,\varphi] +N\int dy \Gamma^{(2)} (x,y)
\int D \Phi \Phi (y) F[\Phi,\varphi]+
 \\
\Ds+\frac{1}{2!}N\int dy_1 dy_2
\Gamma^{(3)} (x,y_1,y_2)\int D \Phi \Phi(y_1)\Phi(y_2) F[\Phi,\varphi]+ \\
\Ds+\frac{1}{3!}N\int dy_1 dy_2 dy_3 \Gamma^{(4)} (x,y_1,y_2,y_3)
\int D \Phi \Phi(y_1)\Phi(y_2)\Phi(y_3) F[\Phi,\varphi]+\cdots .
\lab{e13} \ea\ee

Due to the invariance of the measure $DJ$ in \re{e9} with respect to the
translation
$J\longrightarrow J+\varepsilon$
where $\varepsilon$ is well diminishing function
\ci{vas},
we have
\be 0=N\int D (\frac{1}{\hbar} J)
\left(\frac{\delta W}{\delta J(x)}-\varphi(x)\right)\exp\left(\frac{i}
{\hbar}W[J]-
\frac{i}{\hbar}\int dx J(x)\varphi (x) \right),\lab{e14} \ee

or, in terms of $\Phi$ and $\Gamma[\Phi]$
\be N\int D \Phi \Phi (x) F[\Phi,\varphi]=\varphi (x)
\exp\left(\frac{i}{\hbar}S[\varphi]\right).\lab{e15} \ee

Using such a technique we can obtain the following relations
\be\ba\Ds N\int D \Phi \Phi(y_1)\cdots\Phi(y_3) F[\Phi,\varphi]=
\varphi (y_1) \cdots \varphi (y_3)
\exp\left(\frac{i}{\hbar}S[\varphi]\right)+ \nonumber \\
\Ds+i\hbar N\int D \Phi  \left\{ \left[-\frac{\delta ^2 \Gamma}{\delta
\Phi (y_1)
\delta\Phi (y_2)}\right]^{-1} \Phi (y_3) + \cdots  \right\}  F[\Phi,\varphi]
+O(\hbar ^2),\lab{e16} \ea \ee
\be
\Ds\ba N\int D \Phi \Phi(y_1)\cdots\Phi(y_5) F[\Phi,\varphi]=
\varphi (y_1) \cdots \varphi (y_5)
\Ds\exp\left(\frac{i}{\hbar}S[\varphi]\right)+  \\
\Ds i\hbar N\int D \Phi  \left\{ \left[-\frac{\delta ^2 \Gamma}{\delta\Phi
(y_1)
\delta\Phi (y_2)}\right]^{(-1)} \Phi (y_3)\cdots\Phi(y_5) + \cdots
\right\}  F[\Phi,\varphi]+O(\hbar ^2) \lab{e17} \ea\ee
(the dots in the figure brackets in both expressions means
the terms with permutations of arguments).

The expression like $\left(\delta^2\Gamma / \delta\Phi^2\right)^{-1}$
in the functional integrals must be expanded in series in powers of $\Phi$.
Then there will be the terms like $\hbar\int D \Phi \Phi \cdots \Phi F$
in the expressions \re{e16} and \re{e17}.
Ones may be replased by the terms
 $\hbar\varphi\cdots\varphi\exp\left(\frac{i}{\hbar}S[\varphi]\right)$.
This substitution gives the mistake of the order $\hbar^2$.

The classical action for scalar $\lambda\varphi^4$ model
at $4$-dimensional space-time is presented as follows
\be S[\varphi]=-\frac{1}{2}\int d^{4}x d^{4}y \varphi(x)K(x-y)\varphi(y)-
\frac{\lambda}{4!}\int d^{4}x \varphi^4(x),\lab{e18} \ee
where
$$ K(x-y)=(\partial^2+m^2)\delta^{4}(x-y),$$
and $m$ and $\lambda$ are bare quantities.
Thus we have
\be \frac{\delta S[\varphi]}{\delta \varphi (x)}=-\int dy K(x-y)\varphi(y)-
\frac{\lambda}{3!} \varphi^3 (x).\lab{e19} \ee
Further we shall omit index $4$ in all the integrals.

Substituting \re{e15}-\re{e17} and \re{e19} into \re{e13} and equating the
expressions in the fronts of the
same powers of $\varphi$ at both sides we get (up to the first order
in $\hbar$)
\be\ba\Ds -(\partial^2+m^2)\delta(x-y)=\Gamma^{(2)}(x,y)-\\
\Ds-\frac{i\hbar}{2} \int dy_1 dy_2 \Gamma^{(4)}(x,y_1 ,y_2 ,y)
\left[ \Gamma^{(2)}(y_1 ,y_2 )\right]^{-1},\lab{e20} \ea\ee
\be\ba\Ds -\lambda\delta(x-y_1)\delta(x-y_2)\delta(x-y_3)=
\Gamma^{(4)}(x,y_1 ,y_2 ,y_3 )+ \\
\Ds+\frac{3}{2!}i\hbar\int dz_1 dz_2 du_1 du_2
\Gamma^{(4)}(x,z_1 ,z_2 ,y_3 )
\left[ \Gamma^{(2)}(z_1 ,u_1 )\right]^{-1}
\left[ \Gamma^{(2)}(z_2 ,u_2 )\right]^{-1} \times \\
\Ds \Gamma^{(4)}(y_1 ,y_2 ,u_1 ,u_2)
-\frac{3!}{5!}10i\hbar \int dz_1 dz_2 \Gamma^{(6)}(x,z_1 ,z_2 ,y_1 ,y_2 ,y_3)
\left[ \Gamma^{(2)}(z_1 ,z_2 )\right]^{-1} \lab{e21} \ea\ee
\be\ba\Ds 0=\frac{1}{5!}\Gamma^{(6)}(x,y_1 ,y_2 ,y_3 ,y_4 ,y_5)+
      \frac{i\hbar}{24} \int dz_1 dz_2 du_1 du_2
\Gamma^{(6)}(x,z_1 ,z_2 ,y_3 ,y_4 ,y_5) \\
\Ds\times \left[ \Gamma^{(2)}(z_1 ,u_1 )\right]^{-1}
\left[ \Gamma^{(2)}(z_2 ,u_2 )\right]^{-1}\Gamma^{(4)}(y_1 ,y_2 ,u_1 ,u_2)
+\cdots.\lab{e22} \ea\ee
(in \re{e22} we have written down all the terms of the order of $\hbar^0$
but not all terms of the order of $\hbar$).
Due to the absence of terms with the odd powers of $\varphi$ in $L_{int}$
the Green's
functions $\Gamma^{(n)}$ with odd n are absent too,
i.e. $\Gamma^{(1)}=\Gamma^{(3)}=\cdots=0$.

We can rewrite the three latter expressions in more compact form
(the sense of these condensed notations becomes clear from the
comparison the old and the new expressions)
\be -(\partial^2+m^2)\delta_{ij}=\Gamma^{(2)}_{ij}-
\frac{i\hbar}{2}  \Gamma^{(4)}_{iklj}
\left[ \Gamma^{(2)}\right]^{-1}_{kl}, \ee
\be\ba\Ds -\lambda\delta_{ij}\delta_{ik}\delta_{il}=
\Gamma^{(4)}_{ijkl}+
\frac{3}{2!}i\hbar
\Gamma^{(4)}_{impl}
\left[ \Gamma^{(2)}\right]^{-1}_{mn}
\left[ \Gamma^{(2)}\right]^{-1}_{pr}  \Gamma^{(4)}_{jknr}  -\\
\Ds-\frac{3!}{5!}10i\hbar  \Gamma^{(6)}_{imnjkl}
\left[ \Gamma^{(2)}\right]^{-1}_{mn}  \ea\ee
\be 0 = \frac{1}{5!}\Gamma^{(6)}_{ijklmn}+
      \frac{i\hbar}{24} \Gamma^{(6)}_{iprlmn}
 \left[ \Gamma^{(2)}\right]^{-1}_{ps}
\left[ \Gamma^{(2)}\right]^{-1}_{rt}\Gamma^{(4)}_{jkst}+\cdots.\ee
If we use the similar notations for $S$
\be S[\varphi] = \sum A^{m}\varphi^{m} \ee
then the general form of our expansions will be
\be \ba \Ds A^{m} = \sum_{\beta =0}^{\infty}
\sum_{\sum n_{i}\alpha_{i} - 2\beta = m} s \hbar^{l}
\int \left( \prod_{n_{i}}\left[ \Gamma^{(n_{i})}\right]^{\alpha_{i}} \right)
\left[ \Gamma^{(2)}\right]^{-\beta} (d^{4} x)^{2\beta} \lab{main},\ea\ee
where $n_{i} > 2$, $\alpha_{i} > 0$;
the symmetry factor $s$ and the index $l$ can be found directly from
the expression \re{e13}.

So the "bare" quantities are expressed
in terms of the total vertex Green's functions.
This representation allows us to analyse the
renormalization conditions by the most natural way.
Such a programm will be discussed in the next section.

\section{The renormalization conditions}

The relations \re{e20}, \re{e21}, \re{e22} 
enable us to get the renormalization conditions. 
Let us expand $\Gamma^{(2)}(x,y)$ and $\Gamma^{(4)}(x,y_1,y_2,y_3)$
over $\hbar$:
\be \Gamma^{(2)} = \Gamma^{(2)}_0  +
\hbar\Gamma^{(2)}_1  + \hbar^2 \Gamma^{(2)}_2 + \cdots,\lab{e23} \ee
\be  \Gamma^{(4)} = \Gamma^{(4)}_0 +
\hbar \Gamma^{(4)}_1  + \hbar^2 \Gamma^{(4)}_2 + \cdots.\lab{e24} \ee
Substituting \re{e23} and \re{e24} into \re{e20} and confining ourselves
up to the first order in $\hbar$ we have
\be\ba\Ds -(\partial^2+m^2)\delta(x-y)=\left( \Gamma^{(2)}_0 (x,y)
+\hbar \Gamma^{(2)}_1 (x,y)+\cdots \right)- \\
\Ds-\frac{i\hbar}{2} \int dy_1 dy_2 \left( \Gamma^{(4)}_0 (x,y_1 ,y_2 ,y)+
\cdots \right) \left[ \Gamma^{(2)}_0 (y_1 ,y_2 )+\cdots \right]^{-1}
+O(\hbar^2 ).\lab{e25} \ea\ee

Let us expand the parameters $m$ and $\lambda$ in powers
of $\hbar$:
\be\ba\Ds m^2 =m^2 _0 +\hbar m^2 _1 +\hbar^2 m^2 _2 +\cdots, \nonumber \\
\Ds\lambda=\lambda _0 +\hbar \lambda _1 +\hbar ^2 \lambda _2 +\cdots,
\lab{e26} \ea\ee
here $m^2 _0$, $\lambda _0$, $m^2 _1$, $\lambda _1$ etc.
 are unknown so far; below we shall determine
them. In the all integrals below we shall suppose some
regularization to be introduced.

Comparing \re{e25} and \re{e26} we find
\be \Gamma^{(2)}_0 (x,y)=-(\partial^2+m^2 _0 )\delta(x-y),\lab{e27} \ee

\be \Gamma^{(2)}_1 (x,y)-
\frac{i}{2} \int dy_1 dy_2 \Gamma^{(4)}_0 (x,y_1 ,y_2 ,y)
\left[ \Gamma^{(2)}_0 (y_1 ,y_2 ) \right]^{-1}=-m^2 _1 \delta(x-y) .
\lab{e28} \ee

From \re{e21} we have
\be \Gamma^{(4)}_0 (x,y_1 ,y_2 ,y_3 )=
-\lambda _0 \delta(x-y_1)\delta(x-y_2)\delta(x-y_3),\lab{e29} \ee

which leads to
\be \Gamma^{(2)}_1 (x,y)+
\frac{i}{2}\lambda _0 \delta(x-y) \left[ \Gamma^{(2)}_0 (x,x)
\right] ^{-1}=-m^2 _1 \delta(x-y) .\lab{e30} \ee

From the integral representation of
$\left[\Gamma^{(2)}(x,y)\right]^{-1}$
\be \left[ \Gamma^{(2)}_0 (x,y)\right] ^{-1}= \frac{1}{(2\pi)^4}
\int dk \frac{1}{k^2-m^2 _0}e^{-ik(x-y)},\lab{e31} \ee

and from \re{e30} we get
\be -m^2 _1 \delta(x-y)=\Gamma^{(2)}_1 (x,y)+
\frac{i}{2}\lambda _0 \delta(x-y)\frac{1}{(2\pi)^4}
\int dk \frac{1}{k^2-m^2 _0}.\lab{e32} \ee

Performing Fourier transformation we have
\be -m^2 _1 \delta(p_1+p_2)=\Gamma^{(2)}_1 (p_1,p_2)\delta(p_1+p_2)+
\frac{i\lambda _0}{2}\frac{\delta(p_1+p_2)}{(2\pi)^4}
\int dk \frac{1}{k^2-m^2 _0}.\lab{e33} \ee

From \re{e27} it follows that
\be p^2 -m^2 _0 =\Gamma ^{(2)}_0 (p) .\lab{e34} \ee

Thus, \re{e20} in momentum space (up to the first order in $\hbar$)
has the form:
\be\ba\Ds p^2 -m^2 _0-\hbar m^2 _1 =\Gamma ^{(2)}_0 (p)+\nonumber \\
\Ds+\hbar \left[\Gamma^{(2)}_1 (p)+\frac{i\lambda _0}{2}\frac{1}{(2\pi)^4}
\int dk \frac{1}{k^2-m^2 _0}\right].\lab{e35} \ea\ee

From \re{e21} we get

\be -\lambda_0\delta (p_1 +p_2 +p_3 +p_4)=
\Gamma^{(4)}_0 (p_1,p_2,p_3,p_4)\delta (p_1 +p_2 +p_3 +p_4),\lab{e36} \ee
\be\ba\Ds -\lambda_1\delta (p_1 +p_2 +p_3 +p_4)=
\Gamma^{(4)}_1 (p_1,p_2,p_3,p_4)\delta (p_1 +p_2 +p_3 +p_4)+\nonumber \\
\Ds+\frac{3}{2}i\lambda_0 ^2 \delta (p_1 +p_2 +p_3 +p_4)\frac{1}{(2\pi)^4}
\int dk \frac{1}{k^2-m_0 ^2}\frac{1}{(k-p_1-p_2)^2 -m_0 ^2}, \lab{e37} \ea\ee
and the final form of \re{e21} (up to the first order in $\hbar$)
\be\ba\Ds -\lambda_0-\hbar\lambda_1 = \Gamma^{(4)}_0 (p_1,p_2,p_3,p_4)+
\nonumber \\
\Ds+\hbar\left[\Gamma^{(4)}_1 (p_1,p_2,p_3,p_4)
+\frac{3}{2}i\lambda_0 ^2 \frac{1}{(2\pi)^4}
\int dk \frac{1}{k^2-m_0 ^2}
\frac{1}{(k-p_1-p_2)^2 -m_0 ^2}\right]. \lab{e38} \ea\ee

Let us introduce the demand for the total ("dressed")
vertex functions to be finite.
It's following from this requirement that

(i) $m_0$ and $\lambda_0$ are finite quantities;

(ii) $m _1$ and $\lambda _1$ are infinite ones and they cancel out the
infinite parts of integrals in r.h.s. of \re{e35} and \re{e38};
for example, if we make use of the dimensional regularization then infinite
parts are equal to
$$-\frac{1}{(2\pi)^4}\frac{\lambda _0 m^2 _0}{32 \pi ^2} \frac{1}{4-n}
\;and\; -\frac{1}{(2\pi)^4}\frac{\lambda ^2 _0}{32 \pi ^2} \frac{3}{4-n}.$$
Now $\Gamma_{0}^{(2)}$, $\Gamma_{1}^{(2)}$, $\Gamma_{0}^{(4)}$,
$\Gamma_{1}^{(4)}$ etc. depend only on finite quantities -
$m_{0}$ and $\lambda_{0}$. Acting the same manner we can have $\Gamma^{(n)}$
expressed in term of renormalized mass and coupling constant.

Thus we have
\be \Gamma^{(2)}(p)=p^2 -m_0 ^2 +finite\;parts,\lab{e39} \ee
\be \Gamma^{(4)}(p_1,\cdots,p_4)=-\lambda_0+finite\;parts.\lab{e40} \ee
where "finite parts" mean the finite addition of
the above-mentioned integrals;
they give contributions to the total vertex Green's functions.

As the extraction of infinite parts of any integrals has ambiguities
the finite parts of \re{e39} and \re{e40} are also ambiguous.
Hence, formulas \re{e39} and \re{e40} represent a general form of
renormalization conditions. The finite parts in
\re{e39} and \re{e40} can be specified
only by specifications of $m_0$ and $\lambda_0$.
The requirement that $m_0$ and $\lambda_0$ are physical
mass and coupling constant is equivalent to putting finite
parts in \re{e39} and \re{e40} equal to zero.Putting
\be \Gamma^{(2)}(p=0)=-\overline{m} _0 ^2 ,\lab{e41} \ee
\be \Gamma^{(4)}(p_1=\cdots=p_4=0)=-\overline{\lambda} _0\lab{e42} \ee
we introduce some new constants $\overline{m} _0$ and $\overline{\lambda} _0$
which are related to the previous ones
through finite renormalization \ci{vsh},\ci{iz},\ci{col}.

Let us write down the terms of the order $\hbar ^2$ from \re{e13}
\be\ba\Ds -m^2 _2 \delta(x-y)= \Gamma^{(2)}_2 (x,y) +
\nonumber\\
\Ds+\frac{i}{4}\lambda_0 m^2 _1
\delta (x-y) \int dz \left[\Gamma^{(2)}_0 (x,z)\right]^{-1}
\left[\Gamma^{(2)}_0 (z,x)\right]^{-1}-\nonumber\\
\Ds-\frac{i}{2}\lambda_1 \delta (x-y) \left[\Gamma^{(2)}_0 (x,x)\right]^{-1}-
\nonumber\\
\Ds-\frac{1}{4}\lambda_0 \delta (x-y) \int dz
\left[\Gamma^{(2)}_0 (x,z)\right]^{-1}\left[\Gamma^{(2)}_0 (z,z)\right]^{-1}
\left[\Gamma^{(2)}_0 (z,x)\right]^{-1} -
   \nonumber\\
\Ds-\frac{1}{6}\lambda^2 _0 \left[\Gamma^{(2)}_0 (x,y)\right]^{-3}.\lab{e43}
\ea\ee

Substituting $\left[\Gamma^{(2)}_0 (x,y)\right]^{-1}$ into \re{e43} we get
\be\ba\Ds -m^2 _2 \delta(x-y)= \Gamma^{(2)}_2 (x,y) +\frac{i}{4}\lambda_0 m_1
\delta(x-y) C_1 - \frac{i}{2}\lambda_1\delta(x-y) C_2 -\nonumber\\
\Ds-\frac{1}{4}\lambda_0 \delta(x-y) C_3 -\frac{1}{6}\frac{\lambda_0 ^2 }
{(2\pi)^{12}} \int
\frac{dk_1 dk_2 dk_3 e^{i(k_1 +k_2 +k_3)(x-y)} }
{(k_1 ^2 -m_0 ^2)(k_2 ^2 -m_0 ^2)(k_3 ^2 -m_0 ^2)}
                         ,\lab{e44} \ea\ee
where
$$ C_1 =\frac{1}{(2\pi)^4} \int \frac{dk}{[k^2 -m_0 ^2]^2} ,$$
$$ C_2 =\frac{1}{(2\pi)^4} \int \frac{dk}{k^2 -m_0 ^2} ,$$
$$ C_3 =\frac{1}{(2\pi)^8} \int \frac{dk_1 dk_2}
{[k_1 ^2 -m_0 ^2]^2 (k_2 ^2 -m_0 ^2)}.$$

Performing Fourier transformation and restoring
all the terms of lower orders we have
\be\ba\Ds p^2 -m_0 ^2 -\hbar m_1 ^2 - \hbar^2 m^2 _2 =
\Gamma^{(2)}_0 (p) +  \hbar \left[ \Gamma^{(2)}_1 (p) +
\frac{i}{2}\lambda_0\int \frac{dk}{k^2 -m_0 ^2}\right]+ \nonumber\\+
\Ds\hbar^2 \left[\Gamma^{(2)}_2 (p) +\frac{i}{4}\lambda_0 m_1 ^2 C_1 -
\frac{i}{2}\lambda_1 C_2
-\frac{1}{4}\lambda_0 C_3 -\frac{1}{6}\lambda_0 ^2 I\right],\lab{e45} \ea\ee
where
$$ I=I(p)=\frac{1}{(2\pi)^8} \int
\frac{dk_1 dk_2}{(k_1 ^2 -m_0 ^2)(k_2 ^2 -m_0 ^2)((p-k_1 -k_2)^2 -m_0 ^2)}.$$
and $m_{1}^{2}$ and $\lambda_{1}$ are determined by \re{e35} and \re{e38}.
Therefore we can write
\be\ba\Ds p^2 -m_0 ^2 - \hbar^2 m^2 _2 =
\Gamma^{(2)}_0 (p) + \hbar \Gamma^{(2)}_1 (p) + \hbar^2 \Gamma^{(2)}_2 (p) +
\nonumber\\+
\Ds\hbar^2 \left( -\frac{i}{4}\lambda_0
\frac{1}{(2\pi)^4}\frac{\lambda _0 m^2 _0}{32 \pi ^2} \frac{1}{4-n}
C_1 +\frac{i}{2}
\frac{1}{(2\pi)^4}\frac{\lambda ^2 _0}{32 \pi ^2} \frac{3}{4-n}
C_2 -\frac{1}{4}\lambda_0 C_3 -\frac{1}{6}\lambda_0 ^2 I\right)
\Ds \\ + finite\;const.,\lab{e46} \ea\ee
We can require $m^{2}_{2}$ to be infinite and equal to the constant infinite
parts (including $I(m^{2}_{0})$) of the bracket in rhs. But for the
cancellation of the remaining infinity (which itself depend on $p$) we
must require
\be\ba\Ds (p^2 -m_0 ^2)(1- \frac{d \tilde{\Gamma} ^{(2)}}{d p^2} (m_0 ^2) +
\frac{1}{6}\lambda _0 \hbar ^2 \frac{d I(m_0 ^2)}{d p^2})=0,\lab{e47} \ea\ee
i.e.
\be\ba\Ds \frac{d \tilde{\Gamma} ^{(2)}}{d p^2} (m_0 ^2)= 1+
\frac{1}{6}\lambda _0 \hbar ^2 \frac{d I(m_0 ^2)}{d p^2}=Z_{\varphi},
\lab{e48} \ea\ee
where
\be \tilde{\Gamma} ^{(2)} =
\Gamma^{(2)}_0  + \hbar \Gamma^{(2)}_1  + \hbar^2 \Gamma^{(2)}_2 .
\lab{e49} \ee

If we introduce the renormalized (finite) 2 - points 
vertex $\Gamma^{2} _{R}$
as follows
\be \Gamma^{(2)} = Z_{\varphi} \Gamma^{(2)} _{R} ,\lab{e50} \ee
then we can see that (48) leads to
\be \frac{d \Gamma^{(2)} _{R}}{d p^2} (m_0 ^2)= 1  .\lab{e51} \ee
Expanding \re{e47} in another point $\overline{m}_0^2$ (i.e.,
using the ambiguity in finite
parts in \re{e35} and \re{e38}) we shall have $Z_{\varphi}$ and \re{e51}
defined at the point $\overline{m}_0^2$ .

From now on we have to regard $\Gamma^{(2)}$ in \re{e23} as
$\Gamma^{(2)} _{R}$,
but our formulas \re{e27} - \re{e35} will not change because they were
obtained in the first order in $\hbar$ and the distinction between
$\Gamma^{(2)}$ and $\Gamma^{(2)}_{R}$ appears at least in the second order.
Nevertheless if we wanted to study higher orders of $\hbar$ we should
take into consideration the fact that $\Gamma^{(n)}$ are products of two
series in $\hbar$. On the other hand, according to \re{e47} we can
consider $m^{2}$ to be the product of two types of infinities. Hereby,
in point of fact, we have introduced the wavefunction renormatization.

Now our main purpose is to prove the renormalizability of
our model and investigate the question whether
our renormalization conditions change in higher orders or not.

It is obvious, that if there were no terms like $I(p)$ which lead to the
infinities depending on $p$, then the general form of the conditions
(Eq.(39),(40)) would not change, because all constant infinities
(i.e. terms without depending on $p$) may be cancelled out by the
corresponding parts of the bare parameters $m^2$ and $\lambda$
from \re{e26} in the same manner as it was done in \re{e35},\re{e38}.
Moreover, having chosen the same subtraction point in all the orders of
$\hbar$ we can always obtain conditions like \re{e41},\re{e42}. The only
problem can arise because of the necessity to introduce requirement like
\re{e50}.

For the detailed investigation of this problem we should return to our main
expression \re{main} and rewrite it in momentum representation:

\be \ba \Ds A^{m}_{p} = \sum_{\beta =0}^{\infty}
\sum_{\sum n_{i}\alpha_{i} - 2\beta = m} s \hbar^{l}
\int \left(\frac{d^{4} p}{(2\pi)^{4}}\right)^{2\beta}
\left(
\prod_{n_{i}}\left[ \tilde{\Gamma}^{(n_{i})}_{p}\right]^{\alpha_{i}}
\right)
\left[ ( \tilde{\Gamma}^{(2)}_{p} )^{-1} \right]^{\beta}  \lab{star},\ea\ee
where $\delta$-functions are put into $\tilde{\Gamma}$
\be \tilde{\Gamma}^{(m)}_{p} (p_{1},\ldots,p_{m}) =
\Gamma^{(m)}_{p} (p_{1},\ldots,p_{m}) \delta (\sum_{1}^{m} p_{i}) ,\ee
and integrations must be carried out over all the momenta in all
$( \tilde{\Gamma}^{(2)}_{p} )^{-1}$.
Acting the same manner as it was done in \re{e27}-\re{e38} and
\re{e43}-\re{e45} we can have $\tilde{\Gamma}^{(m)}_{p}$
expressed in term of $( \tilde{\Gamma}^{(2)}_{0} )^{-1}$  and
$\tilde{\Gamma}^{(4)}_{0}$, i.e.
\be \tilde{\Gamma}^{(m)}_{p}  =
\tilde{\Gamma}^{(m)}_{p}
\left[
(\tilde{\Gamma}^{(2)}_{0} )^{-1}, \tilde{\Gamma}^{(4)}_{0}
\right] .\ee

As in our new notations
\be ( \tilde{\Gamma}^{(2)}_{0} )^{-1} (p_{1},p_{2}) =
\frac{\delta (p_{1} + p_{2})}{p_{1}^{2} - m_{0}^{2}} ,\ee
\be \tilde{\Gamma}^{(4)}_{0} (p_{1},\ldots ,p_{4}) =
- \lambda _{0} \delta (p_{1} + \ldots + p_{4}) \ee
we find from \re{star}
\be \ba \Ds A^{m}_{p} = \sum_{\beta =0}^{\infty} K
\int \left(\frac{d^{4} p}{(2\pi)^{4}}\right)^{2\beta}
\prod_{n_{i}}\left[ \tilde{\Gamma}^{(4)}_{0}\right]^{\frac{1}{4}(m+2\beta)}
\left[ ( \tilde{\Gamma}^{(2)}_{0} )^{-1} \right]^{\beta}  \lab{ast},\ea\ee
where $K$ contains both the symmetry factor and $\hbar$.
The representation like \re{ast} i.e. the expansion in series over
$[\Gamma^{(2)}]^{-1}$ is more convenient for analysing the indices of
overall divergency of infinite integrals.

Now we can find the index of divergency $N_{m}$ of the common term
in \re{ast} (we mean, certainly, the index of $p$ in integrand)
\be N_{m} = 4\cdot 2\beta - 2\beta - 4\left( \beta + \frac{1}{4}(m+2\beta)
- 1 \right) = 4 - m .\ee
Here
\newline
$4\cdot 2\beta$ appears from integrals over momenta,
\newline
$-2\beta$ from $( \tilde{\Gamma}^{(2)}_{0} )^{-1}$,
\newline
$- 4\left( \beta + \frac{1}{4}(m+2\beta) - 1 \right)$  from all of the
$\delta$-functions in $( \tilde{\Gamma}^{(2)}_{0} )^{-1}$,
$\tilde{\Gamma}^{(4)}_{0}$ and $- 1$ corresponds to the common
$\delta$-function (providing the conversation of 4-momentum) which must
be took out from the integrand.
Hence, in the expansion of any divergent integral
in series of $p^2$ no terms exept possibly terms in front of $(p^{2})^{0}$
(they have been already discussed) and $(p^{2})^{1}$
are infinite because $N_{m} = 2,0,-2,\ldots$. The terms in front of
$(p^{2})^{1}$ can be always dealt with as well as it was done in
\re{e47} - \re{e50}. Thus, the obtained conditions have the same form
irrespective of the order in $\hbar$ we want to study. In other words,
having proved that the renormalization conditions have the same form in
all orders of $\hbar$ we proved thereby, that our model is
renormalizable.

Let us discuss briefly the unrenormalizable theories. In this case
the infinities stand not only in front of the two first terms in expansion
in $p^2$ but also in front of the higher terms. Therefore our reasons used
for obtaining \re{e39},\re{e40},\re{e51} will not be enough to cancel out
the terms of orders older than $(p^{2})^{1}$. 
In other words, no redifinition
$m^2$ and $\lambda$ and no conditions like \re{e50} will be able to remove
the infinities in the terms of order $(p^{2})^{2}$, $(p^{2})^{3}$ etc.
That is why the renormalization conditions will make no sense any more
as themselves will contain infinities.

So we have studied completely the three renormalization conditions which
are usually postulated to renormalize the $\lambda \varphi ^4$ model
or to obtain the finite effective action for this model.
Although our method is applicable to derive the conditions like
\re{e41}, \re{e42} for any other vertices ($\Gamma^{(6)}$, $\Gamma^{(8)}$ 
etc.) they are scarcely usefull in renormalizations
and we shall not discuss them in details.
We only note that these conditions at $p_{i} = 0$ have the form
\be \Gamma^{(n)}(p_{i}=0) = 0 , n>4 . \ee
This fact can be easily established the same way as it was done
for $\Gamma^{(2)}$ and $\Gamma^{(4)}$.

\bb
\bibitem{edd} Englert F and DeDominicis C 1968 {\sl Nuovo Cim.}
{\bf 53A} p 1007.
\bibitem{da} Amit D {\sl Critical Phenomena and Renormalization Group}.
\bibitem{vas} Vasiliev A N { \sl Functionalnie metodi v kvantovoi teorii
polya i statistike } 1976 (Leningrad: State University Press)  (in Russian).
\bibitem{del} Delbourgo R 1976 {\sl Report on Prog.Phys.} {\bf 39} p 345.
\bibitem{iim} Iliopoulos J, Itzikson C and Martin A 1975 {\sl Rev.Mod.Phys.}
{\bf 47} p 165.
\bibitem{vsh} Vladimirov A and Shirkov D 1979 {\sl Uspekhi Fiz.Nauk }
{\bf 129} p 407.
\bibitem{iz} Itzikson C and Zuber J-B  1980 {\sl Quantum field theory}
(New York: McGraw-Hill).
\bibitem{col} Collins J 1984 {\sl Renormalizations} (Cambrige: University
Press).
\eb

\end{document}